\documentstyle[eqsecnum,aps,multicol,bezier,amsmath,psfig]{revtex}

\begin{document}

\draft

\twocolumn[\hsize\textwidth\columnwidth\hsize\csname@twocolumnfalse\endcsname

\title{Response Function of Coarsening Systems}
\author{Ludovic BERTHIER$^{\ddag,\star}$, Jean-Louis BARRAT$^\star$ and Jorge KURCHAN$^\ddag$}

\address{$^\ddag$Laboratoire de Physique \\ ENS-Lyon and CNRS, F-69364, Lyon Cedex 07, France }

\address{$^\star$D\'epartement de Physique des Mat\'eriaux \\ Universit\'e Claude 
Bernard and CNRS, F-69622 Villeurbanne Cedex, France}

\date{\today}

\maketitle

\begin{abstract}
The response function of  domain growth processes, and in particular the 
violation  of the fluctuation-dissipation theorem,
 are studied both analytically and numerically. 
In the asymptotic limit of large times, 
the fluctuation-dissipation ratio $X$, which quantifies this violation, 
tends   to one if $C>m^2$ and to zero if $C<m^2$, corresponding to  the fast (`bulk') and  slow (`domain-wall')  responses, respectively. In this paper,
we focus on the 
pre-asymptotic behavior of the domain-wall response. This response is shown to 
scale with the typical domain length $L(t)$
  as $1/L(t)$ for dimension $d>2$, and as 
$\ln (L(t))/L(t)$   for $d=2$. Numerical results confirming this
analysis are presented.

\end{abstract}

\pacs{LPENSL-TH-03/99}


\twocolumn\vskip.5pc]\narrowtext

\section{Introduction}

Domain growth systems are the paradigm of systems that do not reach equilibrium.
Hence, it has been a recurring theme in the field of 
spin and structural glasses
to think these systems as displaying some  form of coarsening~\cite{fisher,tarjus}.
In such non-equilibrium systems, time translation invariance does not hold, and
all time dependent correlation functions and response functions depend
on two times (the time origin corresponding generally to the 
time at which the system has been quenched into
the non-equilibrium state). In domain growth phenomena, 
 an autocorrelation function $C(t,t')$ of the form $C(L(t)/L(t'))$
is usually interpreted as arising from structures whose size grows as $L(t)$.
A similar functional form, however, is also found to  
describe the out of equilibrium dynamics of mean-field models of glasses~\cite{kurchan2}, although there is by construction 
no  length $L$ in such models.
In fact, the difference between both kinds 
of models only becomes manifest when one also considers 
the {\it  response} functions associated to the correlation functions.

Generally speaking, the fluctuation-dissipation
theorem (FDT), which for equilibrium systems
relates the response functions to the correlation functions,
 does not hold in systems that are out of equilibrium. 
The deviations from FDT are conveniently described by introducing
the Fluctuation Dissipation Ratio  $X(t,t')$ defined through
$T R(t,t')= X(t,t') \frac{\partial C}{\partial t'}(t,t')$, 
where $R$ is a response function and $C$ the associated 
correlation function. 
In mean field models of spin glasses, the behavior
of this FDR has been well established, at least in the asymptotic
limit of large times.
 One has $X \sim 1$ for ``fast'' processes ($t \sim t'$, large
$C$), but a value $0<X<1$ for the ``slow'' processes corresponding 
to well-separated times 
~\cite{kurchan1,kurchan2}.  
This observation prompted several workers~\cite{letitia,barrat} to calculate the
large-time response of pure (non-disordered) coarsening models, in order to
quantify the  similarities and differences  with mean field  models
of glasses. In order to make a comparison with the glassy case, one  computes
  the  staggered response to a  spatially random field, to make the perturbation
uncorrelated with the equilibrium pure states (as is the case, for example
 of a uniform field for a spin glass). 

The result is that $X \sim 0$ for all
but the smallest time differences. In other words, the long-time memory of  coarsening systems 
tends to vanish, unlike in mean-field glass models
where it does not since $X>0$ even at long times.
As far as we know, all systems in which two (or only few) phases separate have $X = 0$ at long times,
and this has been proven under certain assumptions~\cite{franz}. 
Physically, this feature can be understood from the fact that for long times
the response will be dominated by the bulk response of the domains
that form during the coarsening process. The response at time  $t$
of a spin to a field 
applied at time $t'$ will be nonzero only if the spin is not swept by a domain
wall between $t$ and $t'$. Other types of response involve the domain walls
themselves, whose density decreases with time, and therefore vanish in the
limit of large times.

From the experimental  point of view, aging experiments~\cite{aging} 
show  that glasses  
such as spin glasses or molecular glasses  {\em do} have long term memory.
The asymptotic nature of experimental results is however, always questionable.
This is even more the case in numerical studies, which for a number of models
(spin glasses,
 structural glasses, kinetic models and  polymers in  disordered  media~\cite{bibli_fdt}) have obtained results 
in qualitative agreement with mean-field theory ($X \neq 0$).
It is therefore a relevant question to study the deviations from FDT
in the pre-asymptotic limit.
 An understanding of this
pre-asymptotic behavior should allow to distinguish between true
long term memory and a slow approach to a vanishing $X$.

In this paper we present such a study for 
  the ferromagnetic coarsening or phase-separation
of pure (non-disordered) systems 
after a quench at time $t=0$ 
 from a homogeneous phase ($T=\infty$) into a two-phase region ($T<T_c$),
with and without local conservation of the order parameter.
In the thermodynamic limit, the equilibrium state, where the two phases are completely  separated is never achieved.
We  confirm  the previous  results for the absence of long term memory
in the response function~\cite{letitia,barrat}, and then study the scaling
of this response  {\it in the pre-asymptotic regime}  (large but finite times).
It turns out that the model-dependence enters  only through the form of the growth law  $L(t)$.

We present the systems and the dynamical quantities under study in Section \ref{definitions}.
The  numerical simulation is described in the Section \ref{simulation}, while the analytic 
study is presented  in the Section \ref{analytic}.

\section{Models and definitions}
\label{definitions}

The systems considered here will be described by a coarse-grained formulation, with a scalar order
 parameter $\phi(\boldsymbol{r},t)$ and a Ginzburg-Landau free energy functional
\begin{equation}
F[\phi] = \int\nolimits d^d r \left[ \frac{1}{2} {|\boldsymbol{\nabla}\phi|}^2 + 
\frac{1}{4}\phi^{4} - \frac{1}{2} \phi^2 -h \phi \right] ,
\label{landau}
\end{equation}
where $h(\boldsymbol{r},t)$ is the field conjugated to $\phi(\boldsymbol{r},t)$.
Experimental situations under consideration are for example the coarsening in a ferromagnet, 
or spinodal decomposition in a binary alloy.

Domain growth processes have been much studied since the early works of Lifshitz, Slyosov 
and Wagner.
Ref.~\cite{bray} is a very complete review on the topic.

If the order parameter is not locally conserved, we have the  Ginzburg-Landau equation 
\begin{equation}
\frac{\partial \phi}{\partial t} = - \frac{\delta F}{\delta \phi} + \eta,
\label{tdgl}
\end{equation}
where $\eta$ is a gaussian markovian noise term satisfying
 $\langle \eta(\boldsymbol{r},t)\rangle = 0$ and 
$\langle\eta(\boldsymbol{r},t) \eta(\boldsymbol{r'},t')\rangle=
 2T\delta(\boldsymbol{r}-\boldsymbol{r'} ) \delta(t-t')$, and $T$ is the temperature.
When the order parameter is conserved, the evolution  is given by the Cahn-Hilliard equation
\begin{equation}
\frac{\partial \phi}{\partial t} = \nabla^2 \left( \frac{\delta F}{\delta \phi} \right) + \eta.
\label{cahn}
\end{equation}
In that case, the thermal noise is characterized by the two moments of
 the gaussian distribution $\langle\eta(\boldsymbol{r},t)\rangle = 0$, 
and $\langle\eta(\boldsymbol{r},t) \eta(\boldsymbol{r'},t')\rangle =
 -2T\delta(t-t') \nabla^2 \delta(\boldsymbol{r}-\boldsymbol{r'})$.

Interesting dynamical quantities in the study of the out of equilibrium
 properties are the autocorrelation function defined by
\begin{equation}
 C(t,t_{w}) \equiv \frac{1}{V} \int\nolimits d^d r  \left\langle 
\phi(\boldsymbol{r},t) \phi(\boldsymbol{r},t_{w}) \right\rangle,
\end{equation}
and the associated response function $R(t,t_{w}) \equiv \langle \delta \phi(t)/ \delta h(tw) \rangle$.
At equilibrium, these two quantities depend on time difference
 $\tau \equiv t-t_w$ only, and 
 are related by the usual fluctuation dissipation theorem
\begin{equation}
R(\tau) = -\frac{1}{T} \frac{\partial C(\tau)}{\partial \tau}.
\end{equation}
Out of equilibrium we write~\cite{kurchan1}:
\begin{equation}
R(t,t_w) = \frac{X(t,t_w)}{T} \frac{\partial C(t,t_w)}{\partial t_w},
\label{gfdt}
\end{equation}
which defines $X(t,t_w)$ as  the fluctuation dissipation ratio.

The strategy for our study is now standard. The quench of the system takes 
place at $t=0$. In order to compute the correlation $C(t,t_w)$ we record
the configurations  of the system evolving at zero  external field, $h=0$
for times  $t>t_w$. The operation is repeated on several samples in order
to improve the statistics.

 The integral of the linear response function 
  $M(t,t_w) \equiv \int_{t_w}^{t} ds \,R(t,s)$ is computed by letting
the system evolve under the influence of  a small
field switched on at  $t_w$, and recording the magnetization at time $t$.  
The field is random in space and stationary~\cite{barrat}. It is drawn
from a gaussian distribution with first moment 
$\overline  {h(\boldsymbol{r})} = 0$ and second
 $\overline{ h(\boldsymbol{r}) h(\boldsymbol{r'})} = {h_0}^2 \delta(\boldsymbol{r}-\boldsymbol{r'})$, respectively.
In the language of magnetic systems, the integrated response
 function is thus the staggered magnetization
\begin{equation}
M(t,t_w)=\frac{1}{{h_0}^2 V} \int\nolimits d^d r \, 
\overline{\langle h(\boldsymbol{r}) \phi (\boldsymbol{r}, t) \rangle}.
\end{equation}

An important property of  the FDR has
 to be emphasized for the present discussion. This property,
 found analytically within mean-field models and  verified  numerically in various glassy 
systems~\cite{kurchan1,bibli_fdt}, is that in the asymptotic regime of
 $t,t_w \rightarrow \infty$,  $X(t,t_w)$  depends on the times only 
through a {\em non singular} function of
correlation function $C(t,t_w)$, that is  $X(t,t_w) \equiv x(C(t,t_w))$.
When this property holds,
 the generalized FDT (\ref{gfdt}) gives the following relation between $M(t,t_w)$ and $C(t,t_w)$:
\begin{equation}
M(t,t_w) = \frac{1}{T} \int_{C(t,t_w)}^{C(t,t)} dC\,x(C).
\end{equation}
In equilibrium systems, $x=1$, so that one has the relation $T M(t,t_w) = C(t,t)-C(t,t_w)$. More generally, in non-equilibrium systems,
 a parametric plot $M(t,t_w)\,
 vs \,\,C(t,t)-C(t,t_w)$ is independent of $t_w$, and allows a direct 
determination of $x(C)$.

In the pre-asymptotic regime, the parametric plot of $M(t,t_w)$ {\it vs} $C(t,t_w)$ (with $t$ as  the parameter)
will generally depend on $t_w$. Interesting information
can nevertheless be extracted from this plot, as will be seen in the
next section. In particular, a constant slope is
indicative of a constant value of $X$, and a zero slope (plateau in $M$) corresponds to a loss of
memory in the response.

A second property of the FDR is that under certain assumptions~\cite{franz}
it happens to coincide   
 with the static Parisi function $x(q)= \int_{0}^{q} dq' \, P(q')$,
where $P(q)$ is is the probability distribution of overlaps between real replicas of the same system.
For the ferromagnetic case, $P(q)$ is trivial and $P(q) = \delta (q-M^2)$, where $M=M(T)$
 is the magnetization.
Therefore, we expect the FDR to be 1 if $1> C > M^2$, and 0 if $ M^2 > C$.

\section{Simulation of a spinodal decomposition}
\label{simulation}

The Monte-Carlo studies of Ref.\cite{barrat} agreed qualitatively with the above
 behavior of the FDR, but to our knowledge no quantitative results are available yet.
It is moreover clear that the asymptotic regime where the parametric plot
 Response/Correlation is supposed to collapse into a master-curve was not 
reached, the plot still conserving a dependence on $t_w$. In this work, we will
be interested in a quantitative study of this pre-asymptotic behavior.

For this purpose, the stochastic partial differential equation (\ref{cahn}) was numerically solved, in order to model a spinodal decomposition.
Both time and space were discretized.
A $1024 \times 1024$ square lattice with periodic boundary conditions was used.
Spatial derivatives are treated using an implicit spectral method.
Time derivatives were approximated using a simple Euler scheme.
No real improvements have been obtained using a second or fourth order stochastic 
Runge-Kutta algorithms.
The following recurrence relation is then obtained in the discretized Fourier space:
\begin{equation}
\begin{aligned}
\phi & (\boldsymbol{k},t_{n+1})  = \left[ \frac{1}{1+(k^4-k^2)\Delta t} \right] \times \\
& \times \Big[ \phi (\boldsymbol{k}, t_{n}) -k^2\Delta
 t (\phi^3 (\boldsymbol{k}, t_{n}) - h(\boldsymbol{k})) + \sqrt{k} \eta \Big].
\end{aligned}
\label{rec}
\end{equation}
After discretization, the noise term $\eta$ is characterized by 
$\langle \eta(\boldsymbol{r_i},t_n) \eta(\boldsymbol{r_j},t_m) 
\rangle = 2 T \Delta t / (\Delta x \Delta y) \delta_{ij} \delta_{nm}$.
The algorithm is the following. Knowing the fields $\phi(\boldsymbol{r},t_n)$ 
and $\phi^3 (\boldsymbol{r},t_n)$, the Fourier transforms 
$\phi(\boldsymbol{k},t_n)$ and $\phi^3 (\boldsymbol{k},t_n)$ are computed.
The recurrence relation (\ref{rec}) is then used to obtain $\phi(\boldsymbol{k},t_{n+1})$.
Fourier transforming again gives $\phi(\boldsymbol{r},t_{n+1})$.

The influence of the parameters $\Delta x$, $\Delta y$, $T$ and $\Delta t$ on 
the numerical integration is discussed in the literature~\cite{rogers}.
We chose $\Delta x = \Delta y =0.5$, in order to get mesh-size independent results.
The thickness of the domain walls in the late stage of the phase separation is
 indeed about $\xi = 1/\sqrt{2}$, where $\xi$ is the correlation length of the model (\ref{landau}). 
Their structure is hence sufficiently well described by the above discretization.

The role of $\Delta t$ is made less crucial by our choice of an implicit algorithm.
The linear stability analysis of our algorithm gives indeed the following results: the
 ``tangential bifurcation''~\cite{rogers}, that is the small $k$ instability, is still present, 
but it is of course physically essential.
On the contrary, the ``subharmonic bifurcation''~\cite{rogers} does not exist any more.
Hence, the only restriction on the time step is the average magnitude of the noise which
 has to be kept small in order to avoid numerical divergences.
We chose then the highest possible value of $\Delta t$ compatible with the temperature $T$.
A small temperature allows a large $\Delta t$, but obliges to 
work with a very small magnetic field when computing the response 
function (see below).  
We chose finally $T=0.1$ and $\Delta t = 0.2$ 
in order to explore a large time range.
For very short times, a spurious behavior (see the caption of the figure 
\ref{fdt}) related to this rather large value of $\Delta t$ can be observed.
We checked however that this deviations vanish when $\Delta t$ is smaller, and
do not affect the long time evolution we are interested in.

\begin{figure}
\begin{center}
\begin{tabular}{cc}
\psfig{file=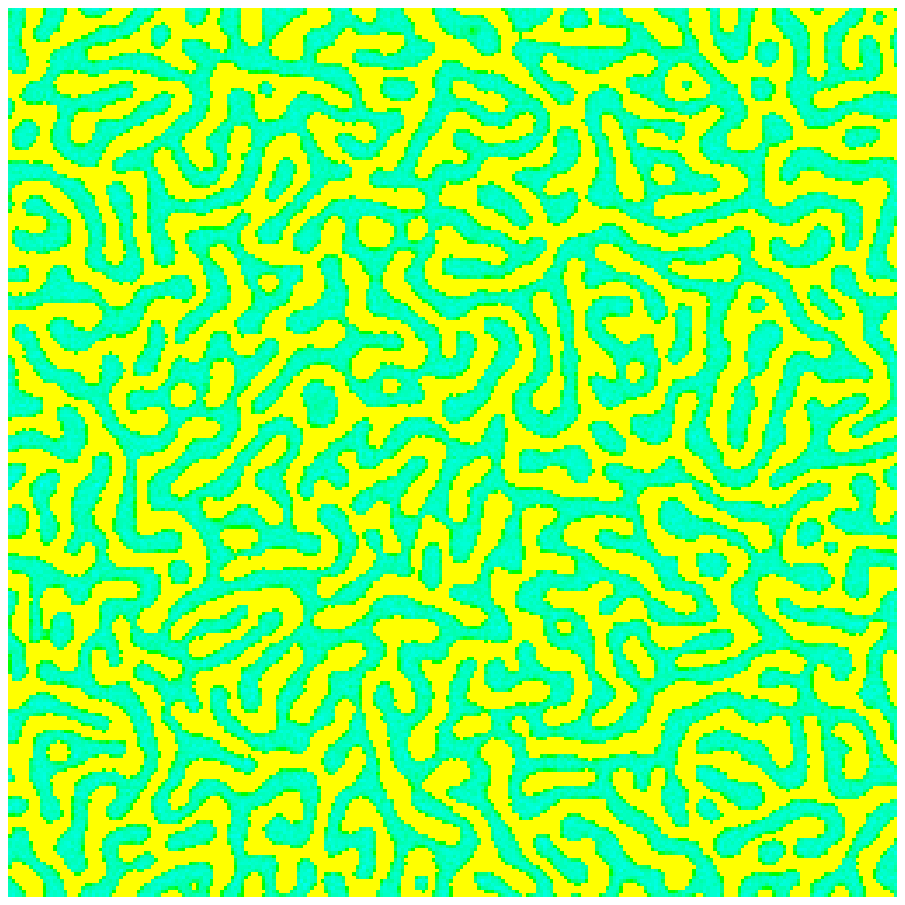,width=3.8cm,height=3.8cm} & \psfig{file=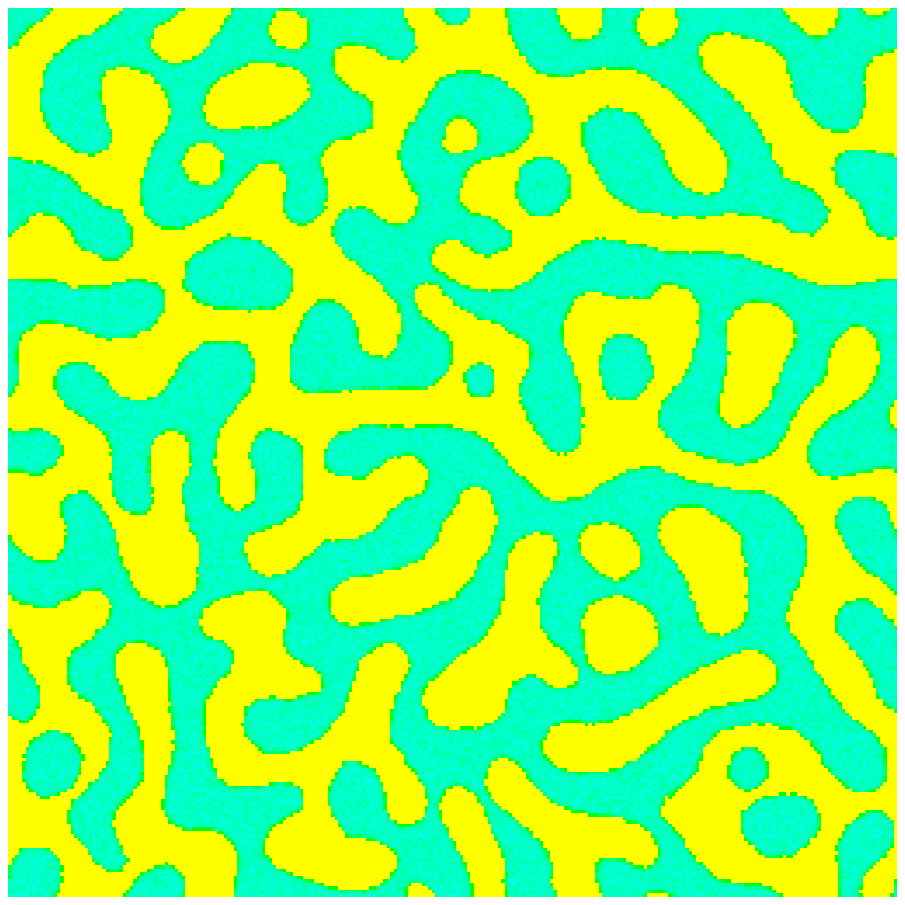,width=3.8cm,height=3.8cm} \\
\psfig{file=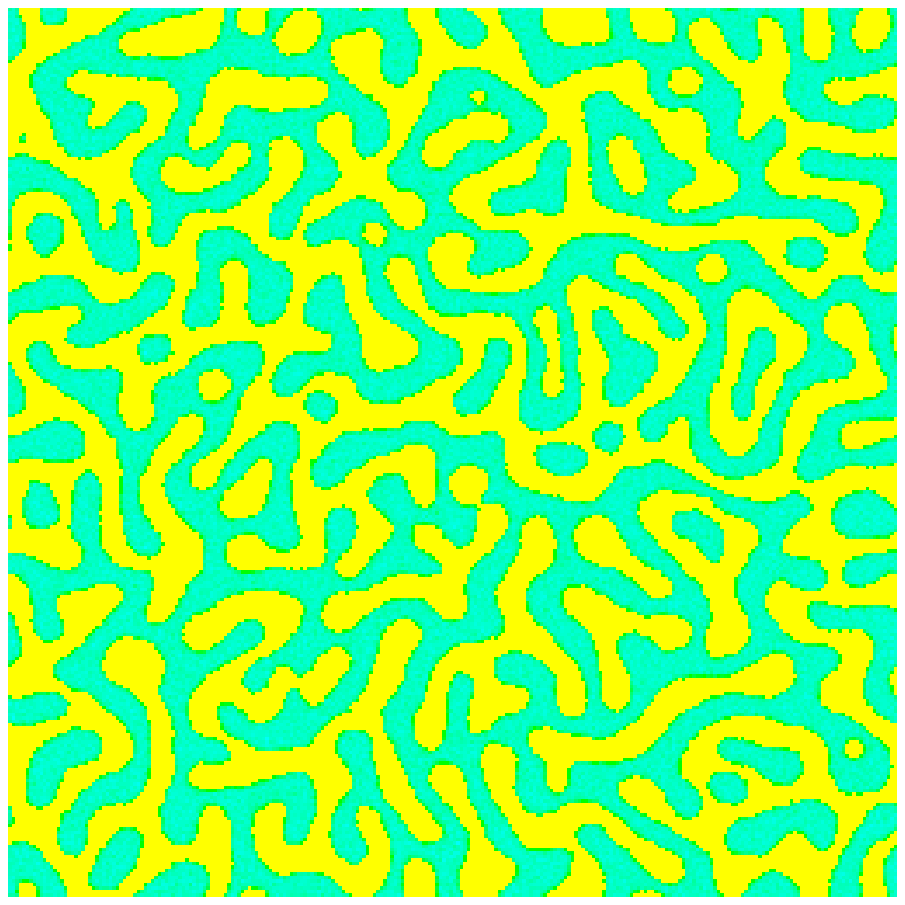,width=3.8cm,height=3.8cm} & \psfig{file=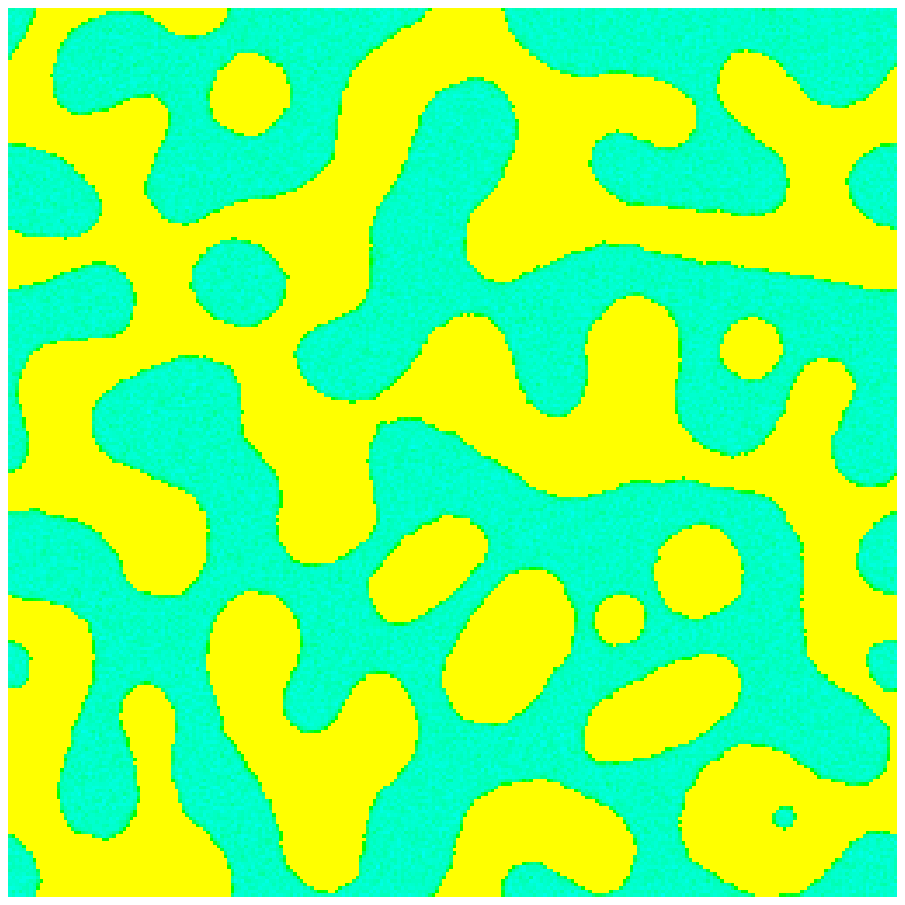,width=3.8cm,height=3.8cm} \\
\end{tabular}
\caption{Field configurations during the simulated phase separation for
 times 317, 1262, 5024, 20100. Each color represents one phase.}
\label{conf}
\end{center}
\end{figure}

Fig.\ref{conf} presents different field configurations during the coarsening process.
Looking at these pictures, it is clear that the coarsening process can be characterized
 by the typical size $L(t)$ of the domains.
Growth laws are well known~\cite{bray} and are $L(t) \sim t^{1/3}$ in the conserved case, 
and $L(t) \sim t^{1/2}$ in the non-conserved case.
The preceding remark has a very interesting consequence which is known as
 {\it the scaling hypothesis}. As $L(t)$ is 
the only physically relevant length scale,
 statistical
 properties of the system are the same 
if we scale all the lengths by the factor $L(t)$~\cite{bray}.

As in experiments, the measure of the domain size is obtained by computing the structure factor 
\begin{equation}
S(\boldsymbol{k},t) \equiv \langle \phi(\boldsymbol{k},t) \phi(- \boldsymbol{k},t) \rangle.
\end{equation} 
The scaling hypothesis implies that it can be written as $S(\boldsymbol{k},t) = L(t)^d g(kL(t))$, 
where $g$ is a scaling function.
A convenient way of obtaining the growth law is to perform a circular average of $S(\boldsymbol{k},t)$ 
and to compute then $\langle k \rangle \equiv \int dk S(k,t)k / \int dk S(k,t)$, which scales as $1/L(t)$.
As in Ref.~\cite{rogers}, we obtained the growth law $L(t) \sim t^{1/3}$, which is valid after a short
 transient period.
The time evolution of the circularly averaged structure factor is depicted in the inset of the figure 
\ref{factor}.
\begin{figure}[t]
\psfig{file=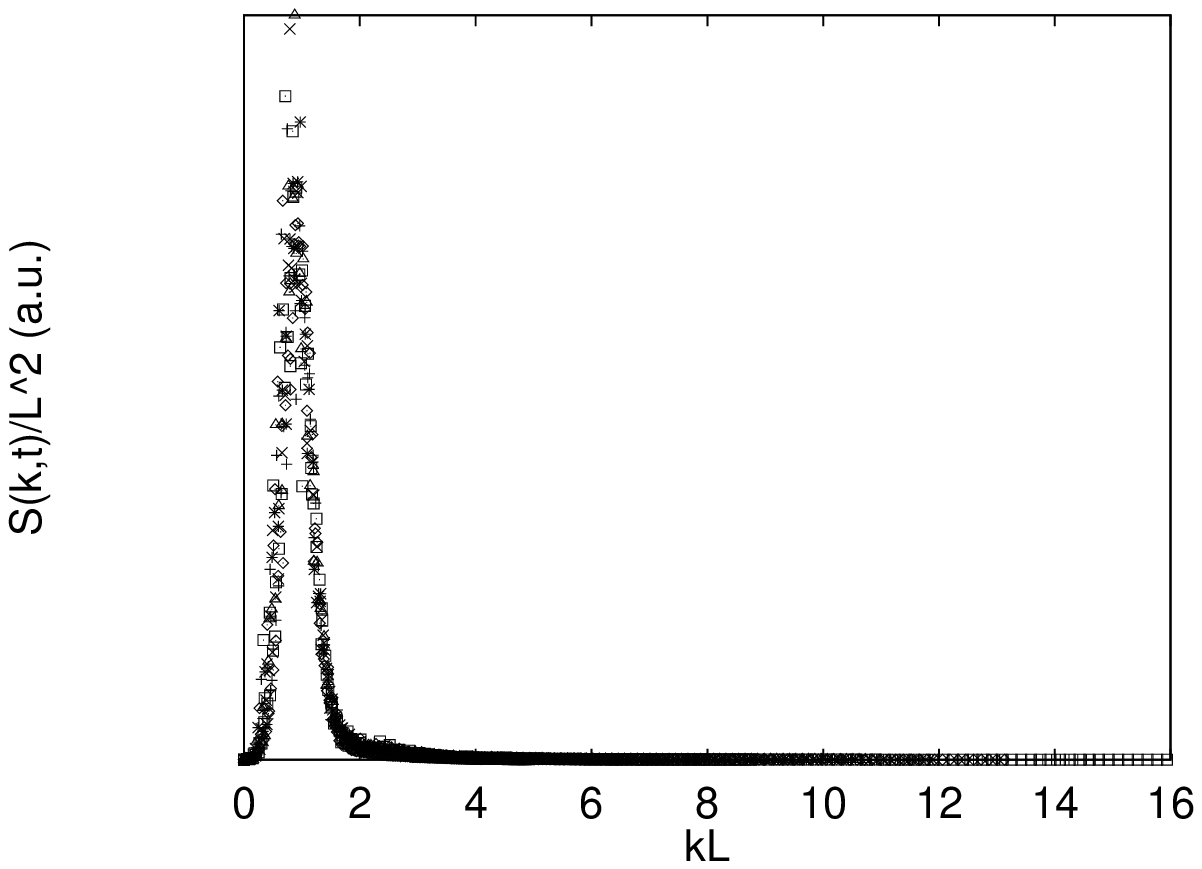,width=8cm,height=6.5cm}
\vspace*{-6.cm}
\hspace*{1.9cm}
\psfig{file=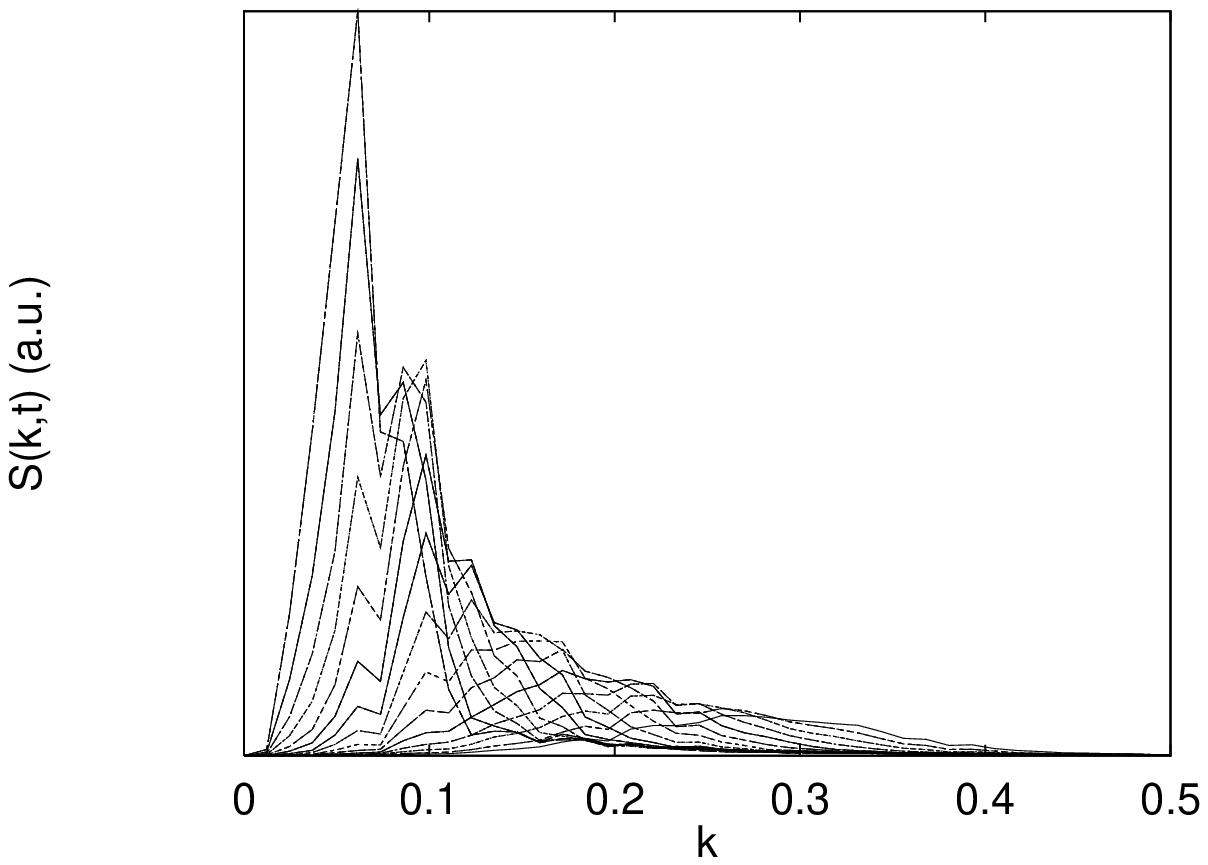,width=5.2cm,height=3.9cm}
\vspace*{2.1cm}
\caption{Inset: circularly averaged structure factor for 15 different times from 317 to 23318. 
The largest time corresponds to the highest maximum. Main picture: the 15 curves collapse 
on a single scaling function.}
\label{factor}
\end{figure}
It has a clear maximum, corresponding to the wave-vector $2\pi /L(t)$.
This maximum shifts towards the small $k$, while its amplitude grows with time.
The scaling hypothesis is verified plotting $S(k,t)/L(t)^2 \,\,vs\,\,kL(t)$, all the curves collapsing on a
 very well defined scaling function $g$.

\begin{figure}
\psfig{file=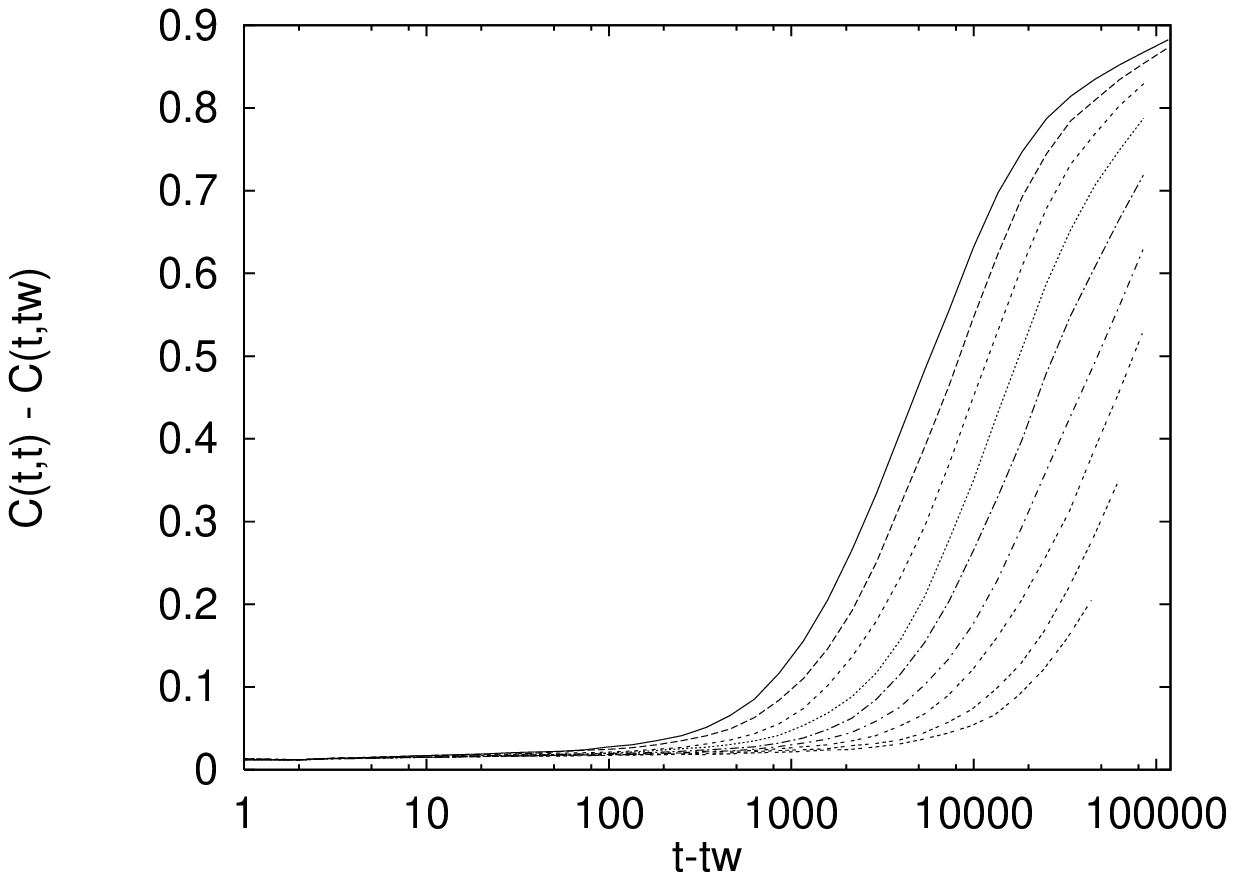,width=8cm,height=6.5cm}
\vspace*{-6.3cm}
\hspace*{0.8cm}
\psfig{file=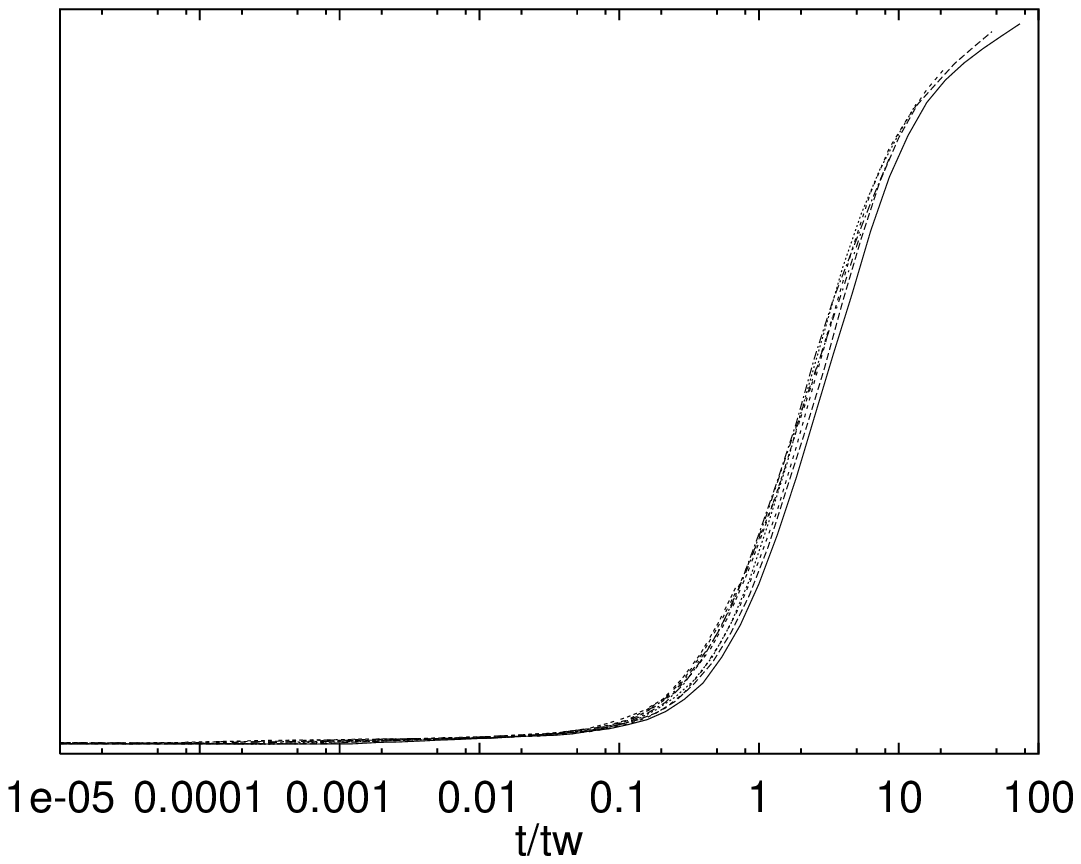,width=4.4cm,height=3.4cm}
\vspace*{3cm}
\caption{Correlation function for waiting times (from left to right)
 $t_w=$ 317, 502, 796, 1262, 2000, 3170, 5024, 7962 and 12619. Inset: the same 
curves as a function of $t/t_w$.}
\label{correlation}
\end{figure}

The autocorrelation functions for various waiting times $t_w$ are shown in the figure~\ref{correlation}.
More precisely, the quantity $C(t,t) - C(t,t_w)$ is computed, in order to avoid normalization problems of
 the correlation functions. As we are working with a soft spin system
rather than a more usual Ising model,
$C(t,t)$ is indeed a slowly varying function. More precisely $C(t,t) \sim C_{eq}-L_0/L(t)$,
where $C_{eq}$ is the equilibrium correlation function in 
a bulk system and  
$L_0/L(t)$, is proportional to the interface density.

Two regimes may be distinguished: for times $t \ll t_w$, the correlation is time translation invariant (TTI),
and for times $t>t_w$, aging is evident, with the TTI breakdown,
 and the correlation falls to 0. This scenario has been called {\it weak ergodicity breaking}~\cite{kurchan1}.
The fluctuation dissipation theorem holds in the former, but is violated
in the latter regime of times.

It is useful to use this behavior to introduce
\begin{equation}
C(t,t_w) = C_{st}(t-t_w) + C_{ag}(t,t_w),
\end{equation}
where $C_{st}$ and $C_{ag}$ describe respectively
 a stationary and an aging part in the correlation.

The scaling hypothesis may be used to predict a scaling form for the aging part of the correlation
 function~\cite{bray}: 
\begin{equation}
C_{ag}(t,t_w)=f \left( \frac{L(t)}{L(t_w)} \right) = f \left( \frac{t}{t_w} \right),
\label{scaling}
\end{equation}
$f$ being a scaling function.
Eq.(\ref{scaling}) retains an explicit dependence on both times $t$ and $t_w$, typical of an aging system.
Such a scaling in the correlation function is called {\it simple aging}.
As shown  in  the inset of figure \ref{correlation}, this scaling form
describes our results extremely well.

In order to complete the study of the fluctuation 
dissipation theorem, we have to compute
 the response function $M(t,t_w)$ to the static
 random field applied between $t_w$ and $t$
 (recall Section \ref{definitions}).
The field amplitude has to be small to obtain a linear response.
The best numerical test we found for this purpose 
is the comparison of the time evolution 
of $\langle k \rangle$ with and without the magnetic field.
When the field is present, the domain walls may be slowed down 
and even pinned if the field is
 too strong, so that the coarsening process is perturbed.
This test is very sensitive, and we worked with small 
field amplitudes (between $h_0 = 0.035$ and $0.09$) to ensure
that the coarsening process was not affected.

\begin{figure}
\psfig{file=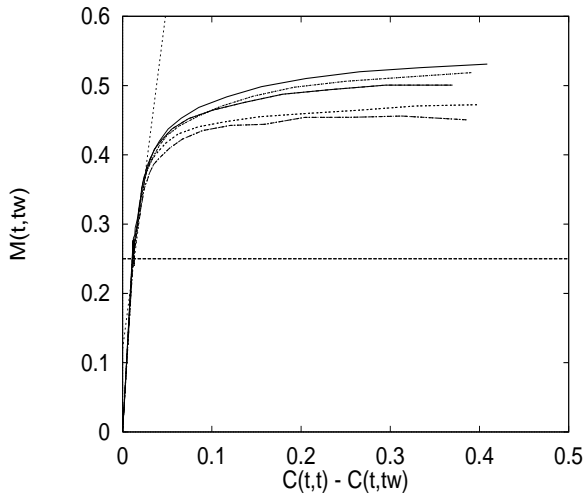,width=8cm,height=6.5cm}
\vspace*{0.2cm}
\caption{Test of the fluctuation dissipation theorem. Waiting times are 
$t_w=$ 317, 502, 796, 2000 and 5024. The horizontal line is the equilibrium 
value of the response and the dashed line is the FDT regime with slope $1/T$.
This line has an intercept which
is slightly positive: this is a time discretization effect, which can be made 
to vanish by reducing $\Delta t$.}
\label{fdt}
\end{figure}

With the correlation and the response functions, the parametric plot
of $M(t,t_w)$ {\it vs} $C(t,t_w)$
 may be built. The data are shown in figure~\ref{fdt}, for various
values of $t_w$, $t$ being the parameter.
The curves are averaged over 14 to 32 realizations of the magnetic field.
They are qualitatively the same as in the previous
 Monte-Carlo simulations~\cite{barrat}, with a first part
 in which  the FDT holds, corresponding to the 
times $t \ll t_w$. In a second part, which corresponds to times 
 $t > t_w$,  the FDT
 is obviously violated, with $M$ having a quasi-horizontal plateau. 
As discussed above, this plateau indicates the loss
of long term memory in the response ($X=0$ at long times),
 consistent with previous expectations.
 
The equilibrium value of the response function  has been
 numerically computed by performing a similar simulation in an  homogeneous system. This value is indicated in figure \ref{fdt} by an horizontal line
(Note that this line can also be determined analytically, as discussed
in the next section).
From the arguments presented in the introduction, it could be expected that
the long time plateau of the integrated response corresponds 
to that of a single domain. It is clearly seen from the data in figure
\ref{fdt}, however, that the approach 
to this asymptotic value is extremely slow.  

As the extra response (i.e. the difference between the plateau value
of $M(t,t_w)$ and the bulk response) can be attributed to the
domain wall response, it is tempting to try to relate this response to the 
domain wall size. 
Between waiting times $t_w=317$ and $t_w=5024$, 
the size of the domains increases
 multiplied by a factor $(5024/317)^{1/3} \sim 2.51$, while the extra response
 is  divided by only $1.36$.
The pre-asymptotic behavior of the FDR seems then to be related to $L(t_w)$ only
 through a non-trivial relation, which we explicitly discuss in the following section.
 
\section{Analytical study of the fluctuation dissipation ratio}
\label{analytic}

In order to study the fluctuation dissipation theorem, we have to compute separately
 the correlation and the response functions.
Eq.(\ref{scaling}) will be sufficient for the present discussion, as we only need a scaling
 form for these functions.
The response function may also be split into
\begin{equation}
M(t,t_w) = M_{eq}(t-t_w) + M_{ag}(t,t_w),
\end{equation}
exactly as we did for the correlation function.

\begin{figure}
\psfig{file=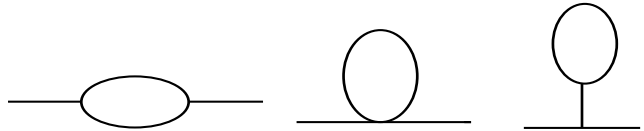,width=8.5cm,height=1.7cm}
\vspace*{0.5cm}
\caption{The three Feynman graphs representing the ``one loop'' (proportional to the temperature $T$)
 corrections to Eq.(\ref{free}).}
\label{feynman}
\end{figure}

We compute first $M_{eq} \equiv \lim_{t-t_w \rightarrow \infty} M_{eq}(t-t_w)$, which is
 in fact the static equilibrium response function of a single domain.
It may be evaluated exactly at $T=0$ (within the gaussian approximation)
and  corrected perturbatively in powers of  $T$.
One easily finds for $T=0$
\begin{equation}
M_{eq} = \int d^d k \frac{1}{k^2 + 1/\xi^2}.
\label{free}
\end{equation}
Recall that we have  $\xi = 1/\sqrt{2}$.
In the simulation, the space is discretized, and
this integral becomes then a discrete sum over the first Brillouin zone.
A numerical evaluation of this sum yields a result in perfect agreement
with the simulation result obtained for an homogeneous system, as described in the previous section. The first temperature corrections
to equation (\ref{free}) can be computed exactly
 (see figure \ref{feynman}), and are indeed found to be negligible at the temperatures we used.

Note that the integral (\ref{free}) is divergent in
 the continuous theory: one has to introduce a spatial cutoff $a$, simulating the underlying lattice
 spacing.
A convenient way of doing this is to multiply the integrand by $\exp (-k^2 a^2)$.
Thus, the equilibrium response function scales with the cutoff as $M_{eq} \sim a^{2-d}$ for $d>2$, and
 as $M_{eq} \sim \ln (a/\xi)$ for $d=2$.

Next, we compute the ``aging part'' of the response function,
which  involves the response of the domain walls.
This can be done using one of the ``approximate theories for scaling functions''~\cite{bray},
 which attempt to give an analytical expression for the scaling function $g$ of the structure 
factor, or equivalently for its Fourier transform.
The spirit of these theories is to replace the field $\phi(\boldsymbol{r},t)$, which at the late 
times of the coarsening process is $\pm1$ outside the domain walls, by an auxiliary field 
$m(\boldsymbol{r},t)$ varying smoothly in space.
This allows to derive evolution equations for $m$ that may -with further approximations- become
 tractable.
This method has already been used to justify analytically the scaling form (\ref{scaling}) of the 
correlation function~\cite{letitia2}, and will be used below for the response function.

Unfortunately, such schemes have not been successfully developed for the 
conserved case.
Our strategy will be then to study the non-conserved case within these approximations,
 and to give physical arguments to extend the validity of our result to the conserved case.    
For this purpose, we generalize a calculation  of Bray in reference ~\cite{braybray} of the response 
 to a uniform magnetic field, to the staggered response to  a random field.
Following Bray, the equation for the auxiliary field $m$ reads (at $T=0$)
\begin{equation}
\frac{\partial m}{\partial t} = \nabla^2 m -n_a n_b \nabla_a \nabla_b m + h | \nabla m |.
\label{bray1}
\end{equation}
The field $h$ depends now on space.
Further simplifications, the validity of which we do not discuss here~\cite{bray,braybray}, are the 
replacements $n_a n_b \rightarrow \delta_{ab} / d$ (circular average), and 
$|\nabla m| \rightarrow \langle(\nabla m)^2 \rangle^{1/2}$.
This computation scheme is near in spirit of the theory of Ohta, Jasnow and Kawasaki~\cite{ojk}.
With these two assumptions, (\ref{bray1}) becomes
\begin{equation}
\frac{\partial m}{\partial t}  = D \nabla^2 m + a(t) h,
\end{equation}
with $D=(d-1)/d$ and $a(t)=\langle(\nabla m)^2 \rangle^{1/2}$.
The solution for $m$ is:
\begin{equation}
m(\boldsymbol{k},t)=m(\boldsymbol{k},0) e^{-k^2 D t} + 
h(\boldsymbol{k}) \int_{t_w}^{t} dt' \, a(t') e^{-k^2 D (t-t')}.
\label{bray3}
\end{equation}
Random initial conditions are conveniently chosen from a gaussian distribution
 with mean zero and variance $\Delta$.
The quantity we want to compute is the staggered magnetization 
$M_{ag}(t,t_w) = \langle h(\boldsymbol{r}) \phi(\boldsymbol{r},t) \rangle / {h_0}^2$.
At late times of the coarsening process, the usual approximation
 $\phi \sim \mathrm{sgn}(m)$ can be made.
Using the fact that the fields $m(\boldsymbol{k},0)$ and $h(\boldsymbol{k})$ 
are gaussian, it is straightforward to obtain the staggered magnetization in term
 of the field $m$ and one gets:
\begin{equation}
M_{ag}(t,t_w) = \sqrt{\frac{2}{\pi}} \frac{\langle h(\boldsymbol{r}) m(\boldsymbol{r}) 
\rangle}{{h_0}^2 \sqrt{\langle m^2 \rangle}}.
\label{bray4}
\end{equation}
Using Eq.(\ref{bray3}), the relations $\langle m^2 \rangle \sim \Delta / (Dt)^{d/2}$ and 
 $a(t) \sim \sqrt{\Delta} / t^{(d+2)/4}$ are also obtained and finally:
\begin{eqnarray}
M_{ag}(t,t_w) &\sim& \int d^d k  \, M(k,t,t_w) 
\label{brayy} \\
M(k,t,t_w) &=&
\int_{t_w}^{t} dt' \, e^{-k^2 D (t-t')} \frac{(Dt)^{d/4}}{(Dt')^{(d+2)/4}}.
\label{bray5}
\end{eqnarray}
This integral over $k$ is divergent for large $k$. As 
  in the equilibrium case we introduce a cutoff length $a$ via a term
 $\exp(-k^2 a^2)$.
The integrals can now be performed and yield
\begin{eqnarray}
\frac{M_{ag}(t,t_w)}  {M_{eq}}    &\sim&
\frac{1}{t_w^{1/2}} F \left( \frac{t}{t_w} \right)  \;\;\;\;\;\;\;\;\; \;\;\;\;d>2, \nonumber \\
 &\sim& \frac{\ln(t_w^{1/2} / a)}{\ln (a/\xi) t_w^{1/2}}  F \left( \frac{t}{t_w} \right)  \;\; d=2. 
\label{scaling2}
\end{eqnarray}
The scaling function $F$ is given by
\begin{equation}
F(\lambda) \equiv \lim_{A \rightarrow 0} \int_1^\lambda \frac{d \lambda'}{A^{2-d}} 
\frac{\lambda^{d/4}}{\lambda^{'(d+2)/4} } \frac{1}{(\lambda-\lambda'+A^2)^{d/2}}
\end{equation} 
Except in dimension $d=2$, the cutoff $a$ disappears if the non-equilibrium response is measured in terms 
of the equilibrium one.

The meaning of this result can be better understood by considering the response associated to
each spatial length scale separately. Defining $M(k,t,t_w)$ as the response
to a sinusoidal perturbation with wave-vector $k$, we can distinguish between
two cases.
 For a wavelength
larger than the domain size, $k \ll 1/L(t_w)$,   we obtain 
\begin{equation}
M(k,t,t_w) \sim t^{d/4} \int_{t_w}^{t} \frac{dt'}{{t'}^{(d+2)/4}} \sim L(t_w)
 G\left( \frac{t}{t_w} \right).
\end{equation} 
This is precisely Eq.(114) in Ref.~\cite{braybray}. The response of long 
wavelength modes grows with time, 
although their effect  becomes negligible because the number of modes 
with  $k \ll 1/L(t_w)$ 
decreases with time.
On the other hand, for short wavelengths,  $k \gg 1/L(t_w)$, the integral can
be  approximated to find 
\begin{equation}
M(k,t,t_w) \sim \frac{1}{k^2} \cdot \frac{1}{t^{1/2}},
\end{equation}
which is in fact a very simple result. The susceptibility of an elastic surface (a flat domain wall)
 when a field with wave-vector $\boldsymbol{k}$ is applied is proportional to $1/k^2$, and 
the density of interfaces is proportional to $1/L(t) \sim 1/t^{1/2}$.

The behavior of the aging part of the response function may now be simply evaluated
 as the sum of two terms.
The first one is the contribution of small wave-vectors $k \ll 1/L(t_w)$.
We  have already shown that for the non-conserved case, it became negligible
as $L \rightarrow \infty$.
We can safely assume that this is a general statement, as the influence 
of the long wavelengths is even smaller in the conserved case ({\it cf} the 
Cahn-Hilliard equation). 
The second one, corresponding to 
wave-vectors $k \gg 1/L(t_w)$ scales as $\int_{1/L}^{1/a} d^d k/k^2 L$.
The long-time response 
\begin{eqnarray}
\frac{M_{ag}(t,t_w)}{M_{eq}}  &\sim&\frac{1}{L(t_w)} 
F\left( \frac{L(t)}{L(t_w)} \right) \;\;\;\;\;\;\;\;\;\;\; \quad d>2, 
\nonumber \\
 &\sim& \frac{\ln(L(t_w) / a)}{\ln(a/\xi)L(t_w)}  F\left( \frac{L(t)}{L(t_w)} \right) \quad d=2,
\label{scaling3}
\end{eqnarray}
is entirely dominated by  the  short wavelengths.
For a non-conserved dynamics, with $L \sim t^{1/2}$, Eq.(\ref{scaling2}) is retrieved.

The result (\ref{scaling3}) is now in a form 
 {\it independent of the dynamics} (conserved or non-conserved order
parameter) of the system.
Hence, for any coarsening system, Eqs.(\ref{scaling})
 and (\ref{scaling3}) give an 
analytical evaluation of the response associated to the domain walls
 ratio in the aging part. As expected, this response vanishes at long times,
so that in the asymptotic 
 regime $t,t_w \rightarrow \infty$, the value $x(C)=0$ is obtained in all dimensions.

\begin{figure}
\psfig{file=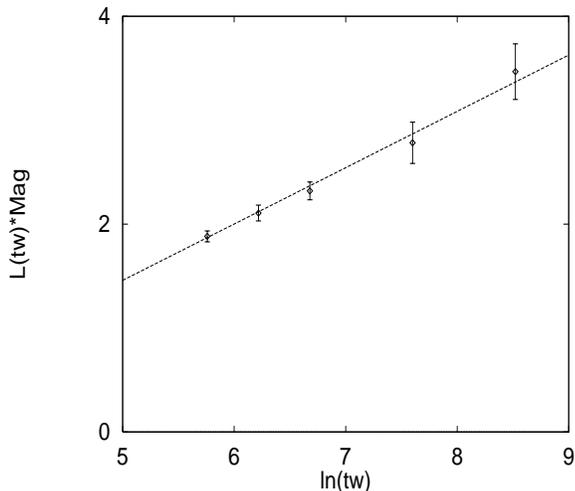,width=8cm,height=6.5cm}
\caption{Test of the scaling (\ref{scaling3}) which predicts a linear
dependence of $L(t_w)M_{ag}$ w.r.t. $\ln (t_w)$. The dashed line fits this 
dependence very  well.}
\label{finaltest}
\end{figure}

The scaling (\ref{scaling3}) is tested in the figure \ref{finaltest}, where we plot 
${t_w}^{1/3} M_{ag}\,\,vs\,\,\ln (t_w)$. Here $M_{ag}$ is defined as the difference between the plateau value obtained in the simulation
and the equilibrium response.
Our data are obviously consistent with the above assumption.
Hence we conclude that the extra response obtained in the simulation
actually corresponds to the domain wall response, and will asymptotically 
vanish. This vanishing, however, is extremely slow, so that we can 
hardly expect to see it in any numerical simulation.

\section{Conclusion}

As mentioned above, 
the important measurable (dynamical) difference between mean-field glass models
and coarsening models is the presence -or absence- of a long term memory
in the response functions.
 In view
 of the relations this has in certain cases with
the Parisi function,  the question as to whether real
 glasses have a value
of $X$ that stays different from, or tends slowly
 to zero is sometimes taken as the
 modern version of the  old ``droplet'' 
versus ``mean-field'' debate of the 80's.
  Experimentally, however, the difference between $X$ 
tending logarithmically to zero 
or staying constant might not be very dramatic.
 What is important, however, is that if one has a law
for the integrated response of the form
\begin{equation}
M_{ag}(t,t_w) \sim A(t_w) F\left( \frac{L(t)}{L(t_w)} \right), 
\end{equation}
then if $A(t)\sim (L(t))^{-1}$  a large aging response 
is necessarily linked to very slow scaling laws $L(t)$. 
The fact that we have found for $d=2$ a relation $A(t) \sim \ln(L(t))/L(t)$
shows that indeed it is possible for the response to fall slower
 than the inverse of the rate of growth $L$,
and one can have relatively large long term memories 
together with rather fast growth laws. 

Another important conclusion of the present study, which confirms earlier
numerical work, is that the domain walls 
can have a  large contribution to the response in the pre-asymptotic regime,
but almost exclusively 
given by  their deformation on relatively
short lengths. This elastic contribution can be considered thermalized,
and its contribution makes longer  the segment of slope $1/T$ in
the parametric plot of the integrated  response versus correlation. 
Apart from that, the plot is flat for smaller values of $C$.
This is still very different from mean-field glass models, in which the
out of equilibrium contribution is `thermalized' at an effective temperature
different from $T$.

\vspace{0.2cm}

This work was supported by the P\^ole
Scientifique de Mod\'elisation Num\'erique at ENS-Lyon.

\end{document}